\documentclass[aps,showpacs,preprintnumbers,amsmath,amssymb,nofootinbibt,twocolumn]{revtex4}
\usepackage{graphicx}
\usepackage{amsmath}
\usepackage{amssymb}
\usepackage{amsfonts}
\usepackage{bm}
\usepackage{color}

\def\be{\begin{equation}}
\def\ee{\end{equation}}
\def\bea{\begin{eqnarray}}
\def\eea{\end{eqnarray}}

\begin{document}


\title{Cosmological evolution of finite temperature Bose-Einstein Condensate dark matter}
\author{Tiberiu Harko}
\email{harko@hkucc.hku.hk} \affiliation{Department of Physics and
Center for Theoretical and Computational Physics, The University
of Hong Kong, Pok Fu Lam Road, Hong Kong, P. R. China}
\author{Gabriela Mocanu}
\email{gabriela.mocanu@ubbcluj.ro} \affiliation{Faculty of Physics, Department of Theoretical and Computational Physics, \\
Babes-Bolyai University, Cluj-Napoca, Romania}

\date{\today}

\begin{abstract}
 Once the temperature of a bosonic gas is smaller than the critical, density dependent, transition temperature, a Bose - Einstein Condensation process  can take place during the cosmological evolution of the Universe. Bose - Einstein Condensates are very strong candidates for dark matter, since they can solve some major issues in observational astrophysics, like, for example, the galactic core/cusp problem. The presence of the dark matter condensates also drastically affects the cosmic history of the Universe.  In the present paper we analyze the effects of the finite dark matter temperature on the cosmological evolution of the Bose-Einstein Condensate dark matter systems. We formulate the basic equations describing the finite temperature condensate, representing a generalized Gross-Pitaevskii equation that takes into account the presence of the thermal cloud in thermodynamic equilibrium with the condensate. The temperature dependent equations of state of the thermal cloud and of the condensate are explicitly obtained in an analytical form. By assuming a flat Friedmann-Robertson-Walker (FRW) geometry, the cosmological evolution of the finite temperature dark matter filled Universe is considered in detail in the framework of a two interacting fluid dark matter model, describing the transition from the initial thermal cloud to the low temperature condensate state. The dynamics  of the cosmological parameters during the finite temperature dominated phase of the dark matter evolution are investigated in detail, and it is shown that the presence of the thermal excitations leads to an overall increase in the expansion rate of the Universe.
\end{abstract}

\maketitle

\section{Introduction}

The Concordance Cosmological Model, usually referred to as the $\Lambda $CDM ($\Lambda $+ Cold Dark Matter) model, has proved to be very successful in explaining cosmological observations across a wide range
of length scales, from the cosmic microwave background (CMB) anisotropy to the Lyman-$\alpha $
forest \cite{PeRa03}. In this model, nonbaryonic collisionless cold dark matter  makes up to
23\% of the total mass content of the Universe.  In the $\Lambda $CDM model dark matter consists of cold neutral weakly interacting massive particles, beyond those existing in the Standard Model of Particle Physics. However, up to now no dark matter candidates have been detected in particle accelerators or in direct and indirect searches. Many particles have been proposed as
possible candidates for dark matter, the most popular ones being  the Weakly Interacting Massive Particles (WIMP) and the axions (for a review of the particle physics aspects of dark matter see \cite{OvWe04}). The
interaction cross section of dark matter particles with normal baryonic matter is assumed to be extremely
small. However,  it is expected to be non-zero, and therefore the direct experimental detection of dark matter particles may be possible by some existing or future detectors \cite{det}.  Superheavy particles, with mass $\geq 10^{10}$ GeV, have also been proposed as dark matter candidates. But in this case  observational results show that these particles must either interact weakly with normal matter, or they must have very heavy masses
above $10^{15}$ GeV \cite{AlBa03}.  Scalar field models, or other long range
coherent fields coupled to gravity have been considered as possible candidates for
galactic dark matter \cite{scal}. The possibility that dark matter could be described by a fluid with non-zero effective pressure was also investigated \cite{pres, pres1}. In particular, it was assumed  that the equation of state of the dark matter halos is polytropic \cite{Sax}. The fit with a polytropic dark halo improves significantly the velocity dispersion profiles. The possibility that dark matter is a mixture of two non-interacting perfect fluids, with different four-velocities and thermodynamic parameters, was proposed recently in \cite{HaLo11}. It has been also suggested that  galactic dynamics of massive test particles can be understood, without considering dark matter, in the context of modified theories of gravity~\cite{dmmodgrav}.

The experimental observation of the Bose-Einstein condensation of dilute alkali gases \cite{exp} represented a major advance in contemporary condensed matter physics. At very low temperatures, all particles in a dilute Bose gas condense to the same quantum ground state, forming a Bose-Einstein Condensate (BEC).  In order for the condensation to occur particles must become correlated with each other so that their wavelengths
overlap, that is, the thermal wavelength $\lambda _{T}$ must satisfy the general condition $\lambda _T>l$,
where $l$ is the mean inter-particles distance. The condensation takes place at a temperature $T<2\pi
\hbar ^{2}/mk_{B}n^{2/3}$, where $m$ is the mass of the particle in the
condensate, $n$ is the particle number density, and $k_{B}$ is Boltzmann's constant
\cite{Da99}. A coherent condensed state always develops if the particle density is high enough, or the temperature is sufficiently low. Quantum degenerate gases have
been created in laboratory by a combination of different laser and evaporative cooling techniques,
opening several new lines of research, at the border of atomic, statistical
and condensed matter physics \cite{Da99}. Recently, the Bose-Einstein condensation of photons has been observed in an optical microcavity \cite{phot}.

The possibility that dark matter, representing a significant amount of the total matter content of our Universe,
is in the form of a Bose-Einstein Condensate was analyzed in detail in \cite{BoHa07}. By introducing the Madelung
representation of the wave function, it follows that dark matter can be described as a non-relativistic, Newtonian Bose-Einstein gravitational condensate gas, whose density and pressure are related by a barotropic equation of state, which, for  a condensate with quartic non-linearity, has the  polytropic index one. The validity of the model was tested by fitting the predicted galactic rotation curves  with a sample of rotation curves of low surface brightness and dwarf galaxies, respectively \cite{BoHa07, jcap, robles}. In all cases a very good agreement was found between the theoretical rotation curves and the observational data. Therefore dark matter halos can be described as an assembly of light individual bosons, occupying the same ground energy state, and acquiring a repulsive interaction.
This interaction prevents gravity from forming the central density cusps. The condensate
particle is light enough to naturally form condensates of very small masses
that later may coalesce.  Different astrophysical properties of condensed dark matter halos have been extensively investigated \cite{BEC}.

The cosmological study of the Bose-Einstein condensate dark matter was initiated recently in \cite{mnr} and \cite{chav}, respectively, with the study of the global cosmological evolution of the cold Bose-Einstein Condensate dark matter, and with the analysis of the perturbations of the
condensate dark matter.  The obtained results show significant differences with respect to the pressureless
dark matter model, considered in the framework of standard cosmology. Therefore the
presence of condensate dark matter could have modified drastically the cosmological
evolution of the early Universe, as well as the large scale structure formation process. The cosmological details of the Bose-Einstein Condensation process have been analyzed for the case of the cold dark matter in \cite{Har}. The evolution of the cosmological inhomogeneities in the condensed dark matter and the observational implications on the Cosmic Microwave Background (CMB) spectra have been considered in \cite{CMB}.

All the mentioned previous researches have been done by assuming a zero temperature condensed dark matter. This assumptions is certainly a very good approximation for the description of dark matter in thermodynamic equilibrium with the cosmic microwave background, and for the analysis of the galactic rotation curves. It is already well established in condensed matter physics that the zero-temperature Gross-Pitaevskii equation gives an excellent quantitative descriptions of the Bose-Einstein condensates for $T\leq 0.5T_{tr}$, where $T_{tr}$ is the Bose-Einstein transition temperature \cite{Gr}. This condition is obviously satisfied by the dark matter halos of the low-redshift galaxies. However, in the early Universe, immediately after the condensation, finite temperature effects could have played an important role in the dark matter dynamics, and significantly influence the cosmological evolution. The study of the finite temperature Bose-Einstein Condensate dark matter was initiated in \cite{HaM}, where the first order temperature corrections to the density profile of the galactic halos and the static properties of
the condensates interacting with a thermal cloud have been obtained.

It is the purpose of the present paper to study the cosmological evolution of the finite temperature
gravitationally self-bound Bose-Einstein dark matter condensates. As a first step in our study, by using the Hartree-Fock-Bogoliubov and Thomas-Fermi approximations, respectively, we obtain the equations of state of the arbitrary finite temperature dark matter condensate interacting with a thermal cloud. A finite temperature condensate can be described in terms of two fluids, the condensate proper, and the thermal excitations. Once the temperature of the system decreases, the number of the thermal excitations drops, and at zero temperature the system consists of the condensate only. Therefore the dynamics of a finite temperature Bose - Einstein -Condensate can be described as a system of two interacting fluids. By using the  specific equations of state of the thermal excitations and of the condensate, the cosmological dynamics of the Universe is considered within a two-fluid dark matter model for a flat Friedmann-Robertson-Walker (FRW) type cosmological model. The physical  parameters of the dark matter are obtained by numerically solving the evolution equations of the system.

The present paper is organized as follows. The basic equations describing finite temperature Bose-Einstein condensates, as well as the properties of the condensate dark matter particles, are reviewed in Section \ref{2}. The equations of state of the finite temperature dark matter, trapped by a gravitational potential, are obtained in Section \ref{3}. The cosmological evolution of the finite temperature condensed dark matter is considered in Section \ref{4}. We discuss and conclude our results in Section \ref{5}.

\section{Finite temperature Bose-Einstein condensate dark matter}\label{2}

In the present Section we briefly review the basic properties of the high temperature Bose-Einstein condensate, and of the dark matter condensed particles. For a detailed discussion of the considered issues and of the derivation of the main results  we refer the reader to the papers and books in \cite{Gr}-{\cite{HFB}.

\subsection{The generalized Gross-Pitaevskii equation and the hydrodynamic representation}

The Heisenberg equation of motion for the quantum field operator $\hat{\Phi}$
describing the dynamics of a Bose-Einstein condensate at arbitrary
temperatures is given by \cite{Za99, HaM, Prouk, Gr}
\begin{eqnarray}
i\hbar \frac{\partial \hat{\Phi}\left(\vec{r},t\right) }{\partial t}&=&\Bigg[
-\frac{\hbar ^{2}}{2m}\Delta +mV_{grav}\left(\vec{r},t\right) +\nonumber\\
&& g^{\prime }\hat{%
\Phi}^{+}\left( \vec{r},t\right) \hat{\Phi}\left(\vec{r},t\right) \Bigg]
\hat{\Phi}\left( \vec{r},t\right) ,  \label{1}
\end{eqnarray}
where $m$ is the mass of the condensed particle, $V_{grav}\left( \vec{r},t%
\right) $ is the gravitational trapping potential, and $g^{\prime }=4\pi l_a\hbar
^{2}/m$, with $l_a$ the $s$-wave scattering length. Eq.~(\ref{1}) is obtained
under the assumption that the interaction potential can be represented as a
zero-range pseudo-potential of strength $g^{\prime }$. By taking the
average of Eq.~(\ref{1}) with respect to a broken symmetry non-equilibrium
ensemble, in which the quantum field operator takes a non-zero expectation
value, we obtain the evolution equation for the
condensate wave-function $\Psi \left(\vec{r},t\right) =\left\langle \hat{%
\Phi}\left( \vec{r},t\right) \right\rangle $. Hence for the exact equation of motion of $\Psi \left( \vec{r},t\right)$ we find
\begin{eqnarray}
i\hbar \frac{\partial \Psi \left( \vec{r},t\right) }{\partial t}&=&\left[ -%
\frac{\hbar ^{2}}{2m}\Delta +mV_{grav}\left( \vec{r},t\right) \right] \Psi
\left( \vec{r},t\right) +\nonumber\\
&& g^{\prime }\left\langle \hat{\Phi}^{+}\left( \vec{r},t
\right) \hat{\Phi}\left( \vec{r},t\right) \hat{\Phi}\left( \vec{r},t%
\right) \right\rangle .  \label{2a}
\end{eqnarray}

By introducing the non-condensate field operator $\tilde{\psi}\left( \vec{r},t
\right) $ so that $\hat{\Phi}\left( \vec{r},t\right) =\Psi \left(\vec{r},t\right) +\tilde{\psi}%
\left( \vec{r},t\right)$,
where the average value of $\tilde{\psi}\left( \vec{r},t\right) $ is zero, $%
\left\langle \tilde{\psi}\left(\vec{r},t\right) \right\rangle =0$, we can separate out the condensate component of the quantum field operator to obtain the equation of motion for $\Psi $ as follows \cite{Za99,HaM,Prouk, Gr}
\begin{eqnarray}
i\hbar \frac{\partial \Psi \left(\vec{r},t\right) }{\partial t}&=&\Bigg[ -%
\frac{\hbar ^{2}}{2m}\Delta +mV_{grav}\left( \vec{r},t\right) +g\rho _{c}\left(
\vec{r},t\right) + \nonumber\\
&&2g\tilde{\rho}\left( \vec{r},t\right) \Bigg] \Psi \left(
\vec{r},t\right) +
g\rho _{\tilde{m}}\left(
\vec{r},t\right)\Psi ^{\ast }\left(
\vec{r},t\right)+\nonumber\\
&& g\rho _{\tilde{\psi}%
^{+} \tilde{\psi} \tilde{\psi}%
}\left(
\vec{r},t\right) ,  \label{6}
\end{eqnarray}
where we have denoted
\be
g=\frac{4\pi l_a\hbar ^{2}}{m^{2}},
\ee
and we have introduced the local condensate mass density
\be
\rho _{c}\left(
\vec{r},t\right) =mn_{c}\left( \vec{r},t\right) =m\left| \Psi \left( \vec{r},t%
\right) \right| ^{2},
\ee
the non-condensate mass density
\be
\tilde{\rho}\left(
\vec{r},t\right) =m\tilde{n}\left(\vec{r},t\right) =m\left\langle \tilde{%
\psi}^{+}\left( \vec{r},t\right) \tilde{\psi}\left( \vec{r},t\right)
\right\rangle ,
\ee
the off-diagonal (anomalous) mass density
\be
\rho _{%
\tilde{m}}\left(
\vec{r},t\right)=m\tilde{m}\left( \vec{r},t\right) =m\left\langle \tilde{\psi}%
\left( \vec{r},t\right) \tilde{\psi}\left( \vec{r},t\right)
\right\rangle ,
\ee
and the three-field correlation function density
\be
\rho _{\tilde{\psi}%
^{+} \tilde{\psi} \tilde{\psi}%
}\left(
\vec{r},t\right)=m\left\langle \tilde{\psi}%
^{+}\left( \vec{r},t\right) \tilde{\psi}\left( \vec{r},t\right) \tilde{\psi}%
\left( \vec{r},t\right) \right\rangle,
\ee
respectively.

In the following we will restrict our analysis to the range of finite
temperatures where the dominant thermal excitations can be approximated as
high energy non-condensed particles moving in a self-consistent Hartree-Fock
mean field, with local energy \cite{Za99, HaM, Prouk, Gr}
\begin{eqnarray}
\bar{\varepsilon}_{p}\left(\vec{r},t\right) &=&\frac{\vec{p}^{2}}{2m}%
+mV_{grav}\left( \vec{r},t\right) +2g\left[ \rho _{c}\left( \vec{r},t\right) +%
\tilde{\rho}\left( \vec{r},t\right) \right] =\nonumber\\
&&\frac{\vec{p}^{2}}{2m}%
+U_{eff}\left( \vec{r},t\right) ,
\end{eqnarray}
where $U_{eff}\left( \vec{r},t\right) =mV_{grav}\left( \vec{r},t\right) +2g\left[
\rho _{c}\left( \vec{r},t\right) +\tilde{\rho}\left( \vec{r},t\right) %
\right] $. Therefore in the present approximation we neglect the mean field effects associated with the anomalous density $\rho _{\tilde{m}}$ and
and with the three-field correlation function $\left\langle \tilde{\psi}^{+}\tilde{%
\psi}\tilde{\psi}\right\rangle $. In the case of dark matter halos with a large number of particles this represents a very good approximation, the contribution of the anomalous density and of the three-field correlation function to the total density being of the order of a few percents \cite{HaM}.

In the thermal cloud the collision between particles forces a
non-equilibrium distribution to evolve to the static Bose-Einstein
distribution $f^{0}\left( \vec{r},\vec{p}\right) $ \cite{Gr}. Hence the particles in
the thermal cloud are in thermodynamic equilibrium among themselves. By
using a single-particle representation spectrum the equilibrium distribution
of the thermal cloud can be written as
\begin{equation}
f^{0}\left(\vec{p}, \vec{r},t\right) =\left[ e^{\beta \bar{\varepsilon}%
_{p}\left( \vec{r},t\right) -\tilde{\mu}}-1\right] ^{-1},
\end{equation}
where $\beta =1/k_{B}T$, with $k_B$ Boltzmann's constant, and $\tilde{\mu}$ is the chemical potential of the
thermal cloud. In order to determine $\tilde{\mu}$ we assume that the condensate and the thermal cloud components are in local diffusive equilibrium with respect to each other. The requirement of a
diffusive equilibrium between the cloud and the condensate imposes
the condition \cite{Za99,HaM,Gr}
\be
\mu _{c}=\tilde{\mu},
\ee
where $\mu _c$ is the chemical potential of the condensate. Therefore $\mu _c$ also determines the static equilibrium distribution of the particles in the thermal cloud.

The equilibrium  density of the thermal excitations is obtained by integrating the
equilibrium Bose - Einstein distribution over the momentum. Thus we obtain \cite{Gr, HaM, Za99,Prouk}
\begin{equation}
\tilde{\rho}\left(  \vec{r},t\right) =\frac{m}{\left(2\pi \hbar \right)^{3}}\int d^{3}\vec{p}%
f^{0}\left( \vec{p}, \vec{r},t\right) =\frac{m}{\lambda _T^{3}}g_{3/2}\left[
z\left( \vec{r},t\right) \right] ,
\end{equation}
where $\lambda _T=$ $\sqrt{2\pi \hbar ^{2}\beta /m}$ is the de Broglie thermal
wavelength, $g_{3/2}(z)$ is a Bose-Einstein function, and the fugacity $%
z\left( \vec{r},t\right) $ is defined as
\begin{equation}
z\left(\vec{r},t\right) =e^{\beta \left[ \tilde{\mu}-U_{eff}\left( \vec{r},t%
\right) \right] }=e^{-\beta g\rho _{c}\left( \vec{r},t\right) }.
\end{equation}

The pressure $\tilde p$ of the thermal excitations can be obtained from the definition \cite{Gr,Za99}
\begin{equation}
\tilde{p}\left( \vec{r},t\right) =\int \frac{d\vec{p}}{\left( 2\pi \hbar
\right) ^{3}}\frac{p^{2}}{3m}f^{0}\left( \vec{p},\vec{r},t\right) ,
\end{equation}
and is given by \cite{Gr,Za99}
\begin{equation}
\tilde{p}\left( \vec{r},t\right) =\frac{1}{\beta \lambda _{T}^{3}}%
g_{5/2}\left[ z\left( \vec{r},t\right)\right] .
\end{equation}

The generalized Gross-Pitaevskii equation can be transformed to a
hydrodynamic form by introducing the Madelung representation of the wave
function as $\Psi \left( t,\vec{r}\right) =\sqrt{\rho _{c}}\exp \left[
\left( i/\hbar \right) S\left( \vec{r},t\right) \right] $. Then by taking into account that in the present approach we neglect the effects of
the mean field associated with the anomalous density  and
the three-field correlation function, it follows
that Eq.~(\ref{6}) is equivalent to the following hydrodynamic type system \cite{Za99, HaM, Gr}
\begin{equation}
\frac{\partial \rho _{c}}{\partial t}+\nabla \cdot \left( \rho _{c}\vec{v}%
_{c}\right) =0,
\end{equation}
\begin{equation}
\frac{\partial S}{\partial t}=-\left( \mu _{c}+\frac{1}{2}m\vec{v}%
_{c}^{2}\right) ,  \label{7}
\end{equation}
where the local velocity of the condensate is given by $\vec{v}_{c}\left(\vec{r},t%
\right) =\left( \hbar /m\right) \nabla S$. The chemical potential
of the condensate is defined as
\begin{equation}
\mu _{c}=-\frac{\hbar ^{2}}{2m}\frac{\Delta \sqrt{\rho _{c}}}{\sqrt{\rho _{c}%
}}+mV_{grav}\left( \vec{r},t\right) +g\rho _{c}\left( \vec{r},t\right) +2g\tilde{%
\rho}\left(\vec{r},t\right) .
\end{equation}

Eq.~(\ref{7}) can be reformulated as the Euler equation of fluid dynamics for the condensate,
\begin{equation}
m\frac{d\vec{v}_{c}}{dt}=m\left[ \frac{\partial \vec{v}_{c}}{\partial t}%
+\left( \vec{v}_{c}\cdot \nabla \right) \vec{v}_{c}\right] =-\nabla \mu _{c}.
\end{equation}

\subsection{Physical properties of the dark matter condensate particle}

The zero-temperature approximation, which gives a very good description of the galactic dark matter halos \cite{BoHa07, jcap},  allows us to make an estimate of the physical properties of the dark matter particle.   From the analysis of the static Bose-Einstein condensate dark matter halos it follows that  the radius $R$ of the condensate dark matter halo is given by $R=\pi
\sqrt{\hbar ^{2}l_a/Gm^{3}}$ \cite{BoHa07}.
The total mass of the condensate dark matter halo $M$ can be obtained as
$M=4\pi ^2\left(\hbar ^2l_a/Gm^3\right)^{3/2}\rho _c=4R^3\rho _{gh}/\pi $, where $\rho _{gh}$ is the central density of the galactic halo, giving for the mean value $<\rho >$ of the condensate density the expression $<\rho >=3\rho _{gh}/\pi ^2$. The dark matter particle mass  in the condensate is given by \cite{BoHa07}
\bea\label{mass}
 m &=&\left( \frac{\pi ^{2}\hbar ^{2}l_a}{GR^{2}}\right) ^{1/3}\approx
6.73\times 10^{-2}\times \nonumber\\
&&\left[ l_a\left( {\rm fm}\right) \right] ^{1/3}%
\left[ R\;{\rm (kpc)}\right] ^{-2/3}\;{\rm eV}.
\eea
For $l_a\approx
1 $ fm and $R\approx 10$ kpc, the typical mass of the condensate particle is of the order of $m\approx 14$
meV. For $l_a\approx 10^{6}$ fm, corresponding
to the values of $l_a$ observed in terrestrial laboratory experiments, $%
m\approx 1.44$ eV.

An important method of observationally obtaining the properties of dark matter is the study of  the collisions between clusters of galaxies, like the bullet cluster (1E 0657-56) and the baby bullet (MACSJ0025-12).
From these studies one can obtain constraints on the physical properties of dark matter, such as its interaction cross-section with baryonic matter, and the dark matter-dark matter self-interaction cross section. If the ratio $\sigma _m=\sigma /m$ of the self-interaction cross section $\sigma =4\pi l_a^2$ and of the dark matter particle mass $m$ is known from observations, then the mass of the dark matter particle in the Bose-Einstein condensate can be obtained from Eq.~(\ref{mass}) as \cite{HaM}
\begin{equation}
m=\left(\frac{\pi ^{3/2}\hbar ^2}{2G}\frac{\sqrt{\sigma _m}}{R^2}\right)^{2/5}.
\end{equation}

By comparing results from X-ray, strong lensing, weak lensing, and optical observations with numerical simulations
of the merging galaxy cluster 1E 0657-56 (the Bullet cluster), an upper limit (68 \% confidence) for $\sigma _m$ of the order of $\sigma _m<1.25\;{\rm cm^2/g}$  was obtained in \cite{Bul}. By adopting for $\sigma _m$ a value of $\sigma _m=1.25\;{\rm cm^2/g}$, we obtain for the mass of the dark matter particle an upper limit of the order
\begin{eqnarray}
m&<&3.1933\times10^{-37}\left(\frac{R}{10\;{\rm kpc}}\right)^{-4/5}\times \nonumber\\
&&\left(\frac{\sigma _m}{1.25\;{\rm cm^2/g}}\right)^{1/5}\;{\rm g}=
0.1791\times\left(\frac{R}{10\;{\rm kpc}}\right)^{-4/5} \times\nonumber\\
&&\left(\frac{\sigma _m}{1.25\;{\rm cm^2/g}}\right)^{1/5}\;{\rm meV}.
\end{eqnarray}
By using this value of the particle mass we can estimate the scattering length $l_a$ as
\be
l_a<\sqrt{\frac{\sigma _m\times m}{4\pi }}=1.7827\times 10^{-19}\;{\rm cm}=1.7827\times 10^{-6}\;{\rm fm}.
\ee
This value of the scattering length $l_a$, obtained from the observations of the Bullet cluster 1E 0657-56, is much smaller than the value of $l_a=10^4-10^6$ fm corresponding to the BEC's obtained in laboratory terrestrial experiments \cite{exp}.

A stronger constraint for $\sigma _m$ was proposed in \cite{Bul1}, so that $\sigma _m\in(0.00335\;{\rm cm^2/g},0.0559\;{\rm cm^2/g})$, giving a dark matter particle mass of the order
\bea
m&\approx &\left(9.516\times 10^{-38}-1.670\times 10^{-37}\right)\left(\frac{R}{10\;{\rm kpc}}\right)^{-4/5}\;{\rm g}=\nonumber\\
&& \left(0.053-0.093\right)\left(\frac{R}{10\;{\rm kpc}}\right)^{-4/5}\;{\rm meV},
\eea
and a scattering length of the order of
\bea
l_a&\approx &\left(5.038-27.255\right)\times 10^{-21}\;{\rm cm}=\nonumber\\
&&\left(5.038-27.255\right)\times 10^{-8}\;{\rm fm}.
\eea

\section{Equation of state of the finite temperature Bose-Einstein condensate}\label{3}

In the present Section we obtain the equation of state of the Bose-Einstein condensate dark matter in thermal equilibrium with the thermal excitations described as a non-condensate cloud. To obtain the equation of state of the global condensate plus thermal excitations cloud we proceed in two steps: first we explicitly construct the equation of state of the thermal excitations, and then we obtain the temperature dependent equation of state of the condensate. The thermodynamic parameters of the condensate plus thermal excitations system can be obtained by adding the densities and pressures of the thermal excitations and of the condensate, respectively.

\subsection{The equation of state of the thermal excitations}

In order to obtain an easy to handle form of the  matter density of the
finite temperature non-condensate particle in thermodynamic equilibrium with
the condensate, we power expand the   $g_{3/2}(z)$  Bose-Einstein function,
so that
\begin{eqnarray}
g_{3/2}\left( e^{-x}\right) &=&2.612-3.544\sqrt{x}%
+1.460x- \nonumber\\
&&0.103x^{2}+0.00424x^{3}+O\left(x^{7/2}\right) .  \label{exp1}
\end{eqnarray}

For $x<1$ Eq.~(\ref{exp1}) approximates the function $g_{3/2}\left(
e^{-x}\right) $ with an error smaller than 1\%. Therefore the density of the
thermal cloud can be represented as
\begin{eqnarray}\label{rhobar}
\tilde{\rho}&=&\frac{m}{\lambda _{T}^{3}}\Bigg[ 2.612-3.544\sqrt{\beta g\rho
_{c}}+1.460\beta g\rho _{c}-\nonumber\\
&&0.103\left( \beta g\rho _{c}\right)
^{2}+0.00424\left( \beta g\rho _{c}\right) ^{3}\Bigg] .
\end{eqnarray}

The previous results can be written in a more transparent form if we introduce the condensation temperature $T_{tr}$, given by \cite{Da99}
\be
T_{tr}=\frac{2\pi \hbar ^2\rho _{tr}^{2/3}}{\zeta ^{2/3}\left(3/2\right)m^{5/3}k_B},
\ee
 where $\zeta (3/2)$ is the Riemann zeta function, and $\rho _{tr}$ is the density of the dark matter at the condensation moment. Then we obtain immediately
\be
\frac{m}{\lambda _T^3}=\frac{\rho _{tr}}{\zeta (3/2)}\left(\frac{T}{T_{tr}}\right)^{3/2},
\ee
and
\be
\beta g\rho _c=2\zeta ^{2/3}(3/2)\frac{l_a}{m^{1/3}\rho _{tr}^{2/3}}\left(\frac{T}{T_{tr}}\right)^{-1}\rho _c,
\ee
respectively.
By introducing the dimensionless condensate density $\theta $, related to
the condensate density $\rho _{c}$ by the relation
\begin{equation}\label{theta}
\rho _{c}=\rho _{tr}^{2/3}\frac{m^{1/3}}{l_a}\theta ,
\end{equation}
we obtain the thermal cloud density as
\begin{eqnarray}\label{dens}
\tilde{\rho}\left( T,\theta \right) &=&\rho _{tr}\left( \frac{T}{T_{tr}}%
\right) ^{3/2}\Bigg[ 1-2.642\left( \frac{T}{T_{tr}}\right) ^{-1/2}\sqrt{\theta }+\nonumber\\
&& 2.120\left( \frac{T}{T_{tr}}\right) ^{-1} \theta
-0.572\left( \frac{T}{T_{tr}}\right)^{-2} \theta ^2+\nonumber\\
&& 0.088\left( \frac{T}{T_{tr}}\right) ^{-3}\theta ^{3}\Bigg] .
\end{eqnarray}

At the initial moment of the condensation $\rho _c=0$, and therefore $\theta \left(T_{tr}\right)=0$. An order of magnitude estimate of the maximum value of $\theta $ can be obtained  by assuming that the density of the condensate is of the same order of magnitude as the initial transition density, $\rho _c\approx \rho _{tr}$, thus giving $\theta _{\max}\approx l_a\rho _{tr}^{1/3}/m^{1/3}$. For $T_{tr}=10^{10}$ K, we obtain $\rho _{tr}=7.262\times 10^{-22}$ g/cm$^3$, and $\theta _{\max}=1.936\times 10^{-15}$. However, due to the incertitude in the numerical values of the physical parameters, we will consider the behavior of the condensed dark matter system for larger values of $\theta _{\max}$. The variation of the thermal cloud density $\tilde{\rho}$ is represented, as a function of $\theta $, and for different values of $T/T_{tr}$, in Fig.~\ref{FIG1}.

\begin{figure}[!ht]
\includegraphics[width=0.98\linewidth]{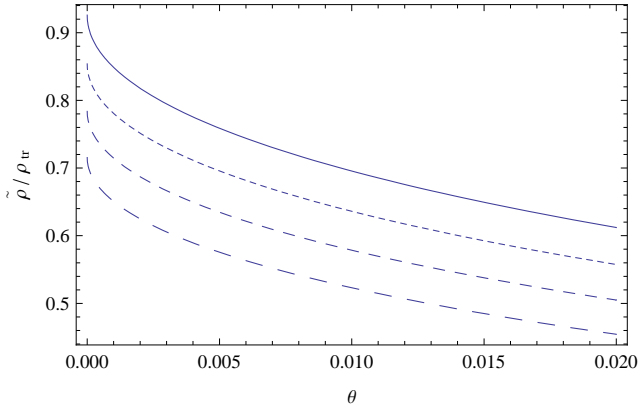}
\caption{The density of the thermal cloud $\tilde{\rho}$ as a function of $\theta $ for different values of $T/T_{tr}$: $T/T_{tr}=0.95$ (solid curve), $T/T_{tr}=0.90$ (dotted curve), $T/T_{tr}=0.85$ (dashed curve) and $T/T_{tr}=0.80$ (long dashed curve), respectively. }
\label{FIG1}
\end{figure}

To obtain the pressure of the non-condensate particles we start with the
expansion
\begin{eqnarray}
g_{5/2}\left( e^{-x}\right)
&=&1.341+2.363x^{3/2}-2.612x-\nonumber\\
&& 0.730x^{2}+0.0346x^{3}+O\left( x^{7/2}\right) ,
\end{eqnarray}
valid for $x<1$, which gives
\begin{eqnarray}
\tilde{p}&=&\frac{k_{B}T}{\lambda _{T}^{3}}\Big[ 1.341+2.363\left( \beta
g\rho _{c}\right) ^{3/2}-2.612\beta g\rho _{c}-\nonumber\\
&& 0.730\left( \beta g\rho
_{c}\right) ^{2}+0.0346\left( \beta g\rho _{c}\right) ^{3}\Big] .
\end{eqnarray}

In terms of the transition temperature we have
\begin{equation}
\frac{k_{B}T}{\lambda _{T}^{3}}=\rho _{tr}\frac{k_{B}T_{tr}}{\zeta \left(
3/2\right) m}\left( \frac{T}{T_{tr}}\right) ^{5/2},
\end{equation}
and thus we obtain  for the non-condensate pressure the expression
\begin{eqnarray}
\tilde{p}\left( T,\theta \right)& =&\rho _{tr}\frac{k_{B}T_{tr}}{m}\left(
\frac{T}{T_{tr}}\right) ^{5/2}\times \nonumber\\
&&\Bigg[ 0.513+6.684\left( \frac{T}{T_{tr}}\right)^{-3/2}\theta
^{3/2}-\nonumber\\
&& 3.793\left( \frac{T}{T_{tr}}\right)^{-1} \theta -4.022\left( \frac{T}{T_{tr}}\right)^{-2}\theta ^{2}+ \nonumber\\
&&0.724\left( \frac{T}{T_{tr}}\right)^{-3}\theta ^{3}\Bigg] .
\end{eqnarray}

The variation of the thermal cloud pressure $\tilde{p}/p_0$, where $p_0=\rho _{tr}k_{B}T_{tr}/m$, is represented, as a function of $\theta $, in Fig.~\ref{FIG2}.

\begin{figure}[!ht]
\includegraphics[width=0.98\linewidth]{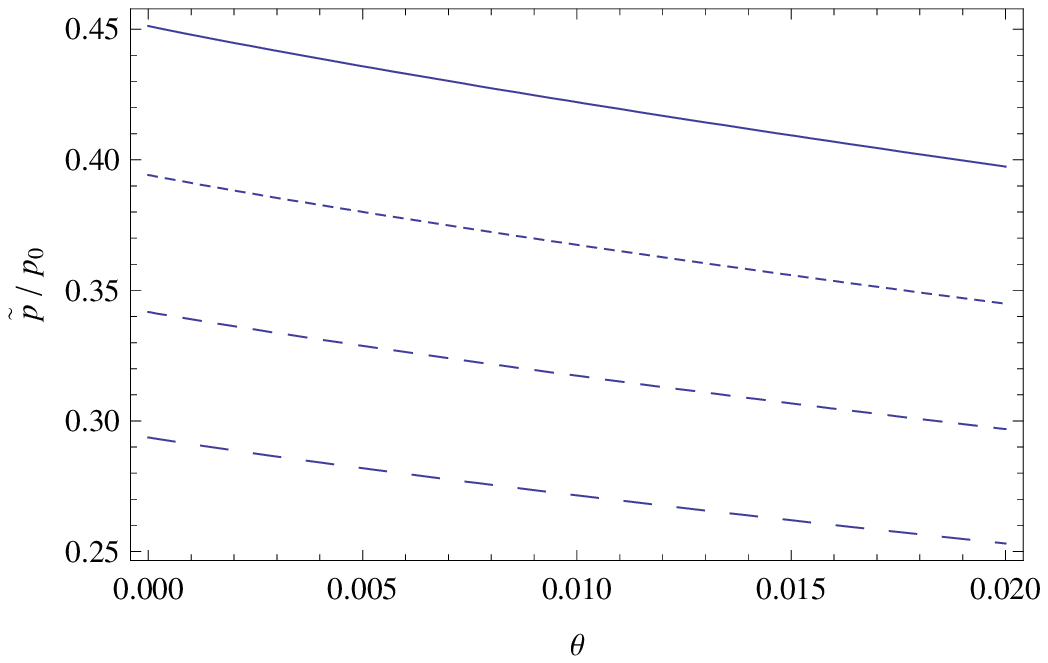}
\caption{The pressure $\tilde{p}/p_0$ of the thermal cloud as a function of $\theta $ for different values of $T/T_{tr}$: $T/T_{tr}=0.95$ (solid curve), $T/T_{tr}=0.90$ (dotted curve), $T/T_{tr}=0.85$ (dashed curve) and $T/T_{tr}=0.80$ (long dashed curve), respectively. }
\label{FIG2}
\end{figure}

The equation of state $\tilde{p}=\tilde{p}\left(\tilde{\rho}\right)$ of the thermal cloud is represented in Fig.~\ref{FIG3}.

\begin{figure}[!ht]
\includegraphics[height=60mm, width=0.98\linewidth]{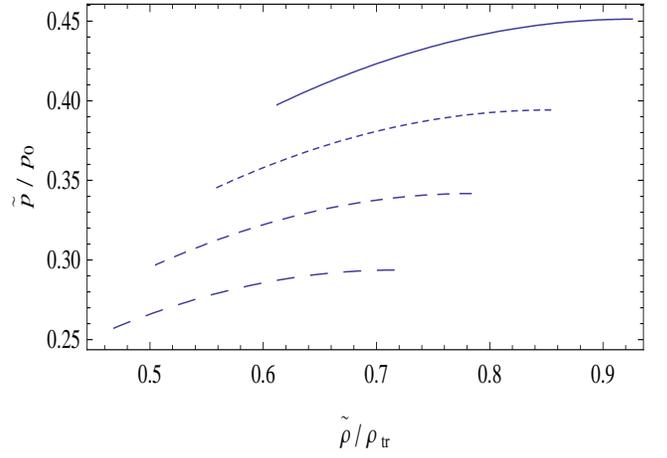}
\caption{The pressure $\tilde{p}/p_0$ of the thermal cloud as a function of $\tilde{\rho}/\rho _{tr}$ for different values of $T/T_{tr}$: $T/T_{tr}=0.95$ (solid curve), $T/T_{tr}=0.90$ (dotted curve), $T/T_{tr}=0.85$ (dashed curve) and $T/T_{tr}=0.80$ (long dashed curve), respectively. }
\label{FIG3}
\end{figure}

The obtained equations of state of the thermal excitations are valid only in the temperature range determined by the condition
\be
k_BT>g\rho _c,
\ee
or, equivalently,
\be
\frac{T}{T_{tr}}>2\zeta ^{2/3}(3/2)\frac{l_a}{m^{1/3}\rho _{tr}^{2/3}}\rho _c.
\ee

\subsection{ The equation of state of the finite temperature condensate}

In order to obtain the equation of state of the condensate we introduce the Thomas-Fermi approximation
for the condensate wave function \cite{Gr} - \cite{HFB}. In the Thomas-Fermi approximation, the
kinetic energy term $-\left( \hbar ^{2}/2m\right) \Delta $ of the condensate
particles is neglected. Hence in this approximation the chemical potential of the condensate is given by \cite{HaM}
\begin{equation}
\mu _{c}=mV_{grav}\left( \vec{r},t\right) +g\rho _{c}\left( \vec{r},t\right) +2g\tilde{%
\rho}\left(\vec{r},t\right),
\end{equation}

Eq.~(\ref{7}) can be reformulated as the Euler equation of fluid dynamics for the condensate,
\begin{equation}\label{euler}
\rho_c\frac{d\vec{v}_{c}}{dt}=-\rho _c\nabla V _{grav}-\frac{g}{m}\rho_c\nabla\left[\rho _{c} +2\tilde{%
\rho}\right].
\end{equation}

By taking into account the identity $\rho _c\nabla \rho _c^n\equiv \nabla \left[n/\left(n+1\right)\right]\rho _c^{n+1}$, and with the use of Eq.~(\ref{rhobar}), it follows that Eq.~(\ref{euler}) of the motion of the Bose-Einstein condensate can be written as
\be
\rho_c\frac{d\vec{v}_{c}}{dt}=-\rho _c\nabla V _{grav}-\nabla p_c,
\ee
where the pressure of the finite temperature Bose-Einstein condensate in thermal equilibrium with a gas of thermal excitations is given by
\bea
p_c\left(T,\rho _c\right)&=&\frac{g}{2m}\rho _c^2-2.362\frac{g^{3/2}}{\lambda _T^3}\left(k_BT\right)^{-1/2}\rho _c^{3/2}+\nonumber\\
&& 1.460\frac{g^2}{\lambda _T^3}\left(k_BT\right)^{-1}\rho _c^2-0.137\frac{g^3}{\lambda _T^3}\times \nonumber\\ &&\left(k_BT\right)^{-2}\rho _c^3+
0.00636\frac{g^4}{\lambda _T^3}\left(k_BT\right)^{-3}\rho _c^4.\nonumber\\
\eea

In terms of the dimensionless density $\theta $, defined by Eq.~(\ref{theta}), we obtain
\bea
p_c\left(T,\theta \right)&=&\rho _{tr}\frac{k_{B}T_{tr}}{m}\Bigg[1.896\frac{m^{1/3}}{\rho _{tr}^{1/3}l_a}\theta ^2
-6.680\left(\frac{T}{T_{tr}}\right)\theta ^{3/2}+\nonumber\\
&&8.043\left(\frac{T}{T_{tr}}\right)^{1/2}\theta ^{2}-
2.863\left(\frac{T}{T_{tr}}\right)^{-1/2}\theta ^{3}+\nonumber\\
&& 0.504\left(\frac{T}{T_{tr}}\right)^{-3/2}\theta ^{4}\Bigg].
\eea
The ratio $p_c/p_0$ for the condensate is determined by the numerical value of the dimensionless parameter
\be
\kappa =\frac{m^{1/3}}{\rho
_{tr}^{1/3}l_{a}}.
\ee

The variation of the condensate pressure with respect to $\theta $ is represented, for $\kappa =50$, Fig.~\ref{FIG4}.

\begin{figure}[!ht]
\includegraphics[height=60mm, width=0.98\linewidth]{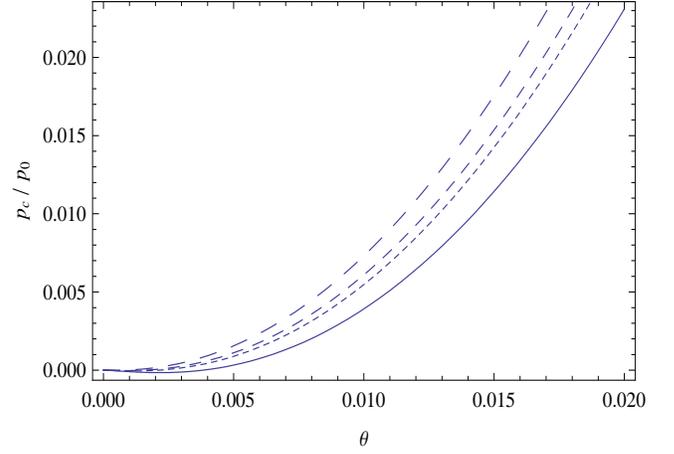}
\caption{The normalized pressure $p_c/p_0$ of the condensate as a function of $\theta $ for $\kappa =50$ and for different values of $T/T_{tr}$: $T/T_{tr}=0.95$ (solid curve), $T/T_{tr}=0.70$ (dotted curve), $T/T_{tr}=0.60$ (dashed curve) and $T/T_{tr}=0.40$ (long dashed curve), respectively. }
\label{FIG4}
\end{figure}

\section{Cosmological evolution of the Universe with finite temperature BEC dark matter}\label{4}

We assume that the space-time geometry of the Universe is the flat Friedmann-Robertson-Walker (FRW) metric, given by
\begin{equation}
ds^{2}=-c^2dt^{2}+a^{2}(t)\left( dx^{2}+dy^{2}+dz^{2}\right),
\label{R6}
\end{equation}
where $a(t)$ is the scale factor describing the cosmological expansion. The Hubble function $H(t)$ is defined as $H=\dot{a}/a$.

For the matter energy-momentum tensor  we consider  the case of the perfect fluid energy-momentum tensor, given by
\begin{equation}
T^{\mu \nu }=(\rho _{tot}c^2+p_{tot})u^{\mu }u^{\nu }+p_{tot}g^{\mu \nu },
\end{equation}
where $\rho _{tot}$ and $p_{tot}$ represent the total energy density and pressure of the Universe.

As for the matter content of the Universe, we assume that it consists of radiation, with energy density $\rho _{rad}$ and pressure $p_{rad}$, pressureless ($p_b=0$) baryonic matter, with energy density $\rho _b$, dark matter, with energy density $\rho _{DM}$, and pressure $p_{DM }$, respectively, and dark energy, described by the cosmological constant $\Lambda $. Hence the total energy density and pressure of the matter in the Universe is given by
\be
\rho_{tot}=\rho _{rad}+\rho _{DM}+\rho _b,
\ee
and
\be
p_{tot}=p_{rad}+p_{DM},
\ee
respectively.
In the following we neglect any possible interaction between these components, by assuming that the energy of each component is individually conserved. Thus, the gravitational field equations, corresponding to the line element given by Eq.~(\ref{R6}), become
\begin{equation}
3\frac{\dot{a}^{2}}{a^{2}} =8\pi G\left(\rho _b+\rho _{rad}+\rho _{DM }\right) +\Lambda ,  \label{dH}
\end{equation}
\begin{equation}
2\frac{\ddot{a}}{a}+\frac{\dot{a}^{2}}{a^{2}} = -\frac{8\pi G}{c^2}\left(p_{rad}+p_{DM}\right)+\Lambda,  \label{dVHi}
\end{equation}
\begin{equation}
\dot{\rho _i}+3\left(\rho _i +\frac{p_i}{c^2}\right)\frac{\dot{a}}{a} =0,\;  i=b, rad, DM . \label{drho}
\end{equation}

From Eq.~(\ref{drho}) it follows that the energy density of the radiation and of the baryonic matter are given by
\be
\rho _{rad}=\frac{\rho_{rad,0}}{\left(a/a_{0}\right)^4},
\ee
and
\be
\rho _b=\frac{\rho_{b,0}}{\left(a/a_{0}\right)^3},
\ee
respectively, where $\rho _{rad,0}$ and $\rho _{b,0}$ are the mass densities corresponding to $a=a_{0}$.

\subsection{Evolution equations of the two-component finite temperature BEC dark matter}

Even if the total energy-density of the condensate-thermal excitations system is conserved, the energies of the condensate and of the thermal cloud are not independently conserved, due to the growth of the condensate through absorption of the particles from the thermal cloud. This process must naturally occur during the expansion and subsequent cooling of the Universe. Moreover, the collisions in the thermal cloud will transfer a particle to the condensate, so that the number $N$ of the condensed particles will increase as $N\rightarrow N+1$,
and there is of course the reverse process, where a collision of a thermal cloud particle with one within the condensate transfers a particle from the condensate into the thermal excitation band, so that $N\rightarrow N-1$. However, in the following we will consider only a simplified model, in which the particles of the thermal cloud turn into the condensate at a rate $1/\tau _{col}$ particles per second, leading to a decrease of a number of particles in the thermal cloud and to the corresponding increase of the number of the condensate particles. Therefore the equations of the energy conservation of the two dark matter components take the form
\be\label{m1}
\frac{d\tilde{\rho }}{dt}+3\frac{\dot{a}}{a}\left(\tilde{\rho }+\frac{\tilde{p}}{c^2}\right)=-\Gamma ^{out}(N),
\ee
and
\be\label{m2}
\frac{d\rho _c}{dt}+3\frac{\dot{a}}{a}\left(\rho _c+\frac{p_c}{c^2}\right)=\Gamma ^{in}(N),
\ee
respectively, where $\Gamma ^{out}(N)$ and $\Gamma ^{in}$ describe the time variation of the density of the two components in the finite temperature dark matter system.
 From Eqs.~(\ref{m1}) and (\ref{m2}) it follows that the total dark matter density $\rho _{DM}=\tilde{\rho }+\rho _c$ satisfies the conservation equation
\be
\frac{d\rho _{DM}}{dt}+3\frac{\dot{a}}{a}\left(\rho _{DM} +\frac{p_{DM}}{c^2}\right)=0,
\ee
where $p_{DM}=\tilde{p}+p_c$ is the total pressure of the condensate-thermal excitations dark matter system.

 The  time rate $\tau _{col}$ at which the excited components turn into the condensate may be approximated by the transition
probability $W(N)$ that can be estimated  by using the quantum kinetic theory, so that $1/\tau_{col}\approx W(N)$ \cite{W}. $W(N)$ gives the rate at which the thermal particles above the condensate energy band enter the condensate due to inter-particle collisions. The main assumption in obtaining $W(N)$ is that the condensate does not readily act
back on the thermal component to change its temperature. This assumption is expected to be valid when one considers
equilibrium or quasi-equilibrium situations. The transition rate of the thermal excitations to the condensed state can be given as \cite{W}
\be\label{m3}
\frac{1}{\tau _{col}}=\frac{4m\left(l_ak_BT\right)^2}{\pi \hbar ^3}e^{2\tilde{\mu }/k_BT}\left[\frac{\mu _c}{k_BT}K_1\left(\frac{\mu _c}{k_BT}\right)\right],
\ee
where $\tilde{\mu }$ and $\mu _c$ are the chemical potentials of the thermal cloud and of the condensate, and $K_1$ is a modified Bessel function. Eq.~(\ref{m3}) can be simplified to the form
\be\label{col}
\frac{1}{\tau _{col}}\approx C_{cor}\frac{4m\left(l_ak_BT\right)^2}{\pi \hbar ^3}=\frac{16\pi\hbar C_{cor}}{\zeta ^{4/3}\left(3/2\right)ml_a^2}\kappa ^{-4}\left(\frac{T}{T_{tr}}\right)^2,
\ee
where the correction factor $C_{cor}$, assumed to be a constant, incorporates the effects of the exponential and of the Bessel function, respectively. The transition rate can be written as
\bea
\frac{1}{\tau _{col}}&\approx &C_{cor}\times 2.0705\times 10^{-28}\times\left(\frac{m}{10^{-37}\;{\rm g}}\right)\times \nonumber\\
&&\left(\frac{l_a}{10^{-20}\;{\rm cm}}\right)^2\left(\frac{T}{{\rm K}}\right)^2\;{\rm s}^{-1}.
\eea

Generally, the source terms in the thermal cloud and continuity equations
can be written as \cite{Gr}
\begin{eqnarray}\label{g1}
\Gamma ^{out} &=&\frac{\sigma \rho _{c}}{\pi m^{3}\left( 2\pi \hbar \right)
^{3}}\int d\vec{p}_{2}d\vec{p}_{3}d\vec{p}_{4}\delta \left( \vec{p}_{c}+\vec{%
p}_{2}-\vec{p}_{3}-\vec{p}_{4}\right) \times  \nonumber\\
&&\delta \left( \varepsilon _{c}+\varepsilon _{2}-\varepsilon
_{3}-\varepsilon _{4}\right) \times f_{2}\left( 1+f_{3}\right) \left(
1+f_{4}\right) ,
\end{eqnarray}
and
\begin{eqnarray}\label{g2}
\Gamma ^{in} &=&\frac{\sigma \rho _{c}}{\pi m^{3}\left( 2\pi \hbar \right)
^{3}}\int d\vec{p}_{2}d\vec{p}_{3}d\vec{p}_{4}\delta \left( \vec{p}_{c}+\vec{%
p}_{3}-\vec{p}_{2}-\vec{p}_{4}\right) \times  \nonumber\\
&&\delta \left( \varepsilon _{c}+\varepsilon _{3}-\varepsilon
_{2}-\varepsilon _{4}\right) \times f_{2}\left( 1+f_{3}\right) f_{4},
\end{eqnarray}
where $\sigma $ is the scattering cross section, $\vec{p}$ and $%
\varepsilon $ denote the particle momentum and energy, and $f_i$, $i=1,2,3,4$ are the Bose-Einstein distribution functions of the thermal particles and of the condensate particles, respectively. The out collision rate
represents the scattering of an incoming thermal particle (2) from the
condensate to produce two thermal atoms (3) and (4). The designation "out" means
that a particle is leaving the condensate. Thus, $\Gamma ^{out}$ is the rate
of increase of the number of thermal particles per unit volume and per unit
time as a result of the collision with a condensate particle. The reverse
process gives the ''in'' collision rate of a thermal particle $\Gamma ^{in}$%
. In the thermodynamic equilibrium state
 \be
 \Gamma ^{out}=\Gamma ^{in}=\Gamma ,
 \ee
and in the following we assume that this condition holds for all times. The equality between the transition rates between the thermal clod and the condensate  also allows us to define the equilibrium collision time $\tau _{col}$, given by Eq.~(\ref{col}). Generally, $\Gamma ^{out}$ and $\Gamma ^{in}$ can be obtained by numerical integration of Eqs.~(\ref{g1}) and (\ref{g2}). However, in order to obtain a semi-analytical model of the cosmological evolution of the finite temperature Bose-Einstein condensate dark matter,
for the energy density transfer rate $\Gamma $ we introduce the simplifying phenomenological assumption that it is proportional to the density of the condensate $\rho _c$, and we write it as
\bea
\Gamma (N)&\propto &\frac{1}{\tau _{col}}\rho _c=4K\frac{m\left(l_ak_BT\right)^2}{\pi \hbar ^3}\rho _c=\nonumber\\
&& \frac{16\pi\hbar K}{\zeta ^{4/3}\left(3/2\right)ml_a^2}\kappa ^{-4}\left(\frac{T}{T_{tr}}\right)^2\rho _c=\nonumber\\
&& 1.472\times 10^{51}\times K\times \left(\frac{m}{10^{-37}\;{\rm g}}\right)^{-1}\times \nonumber\\
&&\left(\frac{l_a}{10^{-20}\;{\rm cm}}\right)^{-2}
\kappa ^{-4}\left(\frac{T}{T_{tr}}\right)^2\rho _c\;\frac{{\rm g}}{{\rm cm^3\;s}},
\eea
where $K$ is a constant. In terms of the dimensionless condensate density $\theta $ we obtain
\bea
\Gamma &=&\frac{16\pi \hbar l_a}{\zeta ^{4/3}(3/2)m^2}K\rho_{tr}^2\left(\frac{T}{T_{tr}}\right)^2\theta =\nonumber\\
&& \frac{16\pi \hbar }{\zeta ^{4/3}(3/2)l_a^5}K\kappa ^{-6}\left(\frac{T}{T_{tr}}\right)^2\theta .
\eea

\subsection{Cosmological dynamics of the finite temperature BEC dark matter}

In the following we denote the "reduced" dimensionless temperature $T/T_{tr}$ by $\chi $,  $\chi =T/T_{tr}$. By taking into account that in the case of the finite temperature
Bose-Einstein condensate dark matter the energy densities $\rho _{c}$ and $\tilde{\rho }$ are
functions of the reduced temperature $\chi $ and of the condensate density $\theta $,   $\rho _{c}=\rho _c\left(\chi ,\theta \right)$,  $\tilde{\rho }=\tilde{\rho }\left( \chi,\theta
\right) $, the energy conservation equation for the dark matter takes the
form
\bea\label{eq1}
&&\partial _{\theta }\tilde{\rho }\left( \chi,\theta \right) \dot{\theta}+\partial
_{\chi}\tilde{\rho }\left( \chi,\theta \right) \dot{\chi}+\nonumber\\
&&3\frac{\dot{a}}{a}\left[ \tilde{\rho }
\left( \chi,\theta \right) +\frac{\tilde{p}\left( \chi,\theta \right) }{c^{2}%
}\right] =-\Gamma ,
\eea
and
\bea\label{eq2}
&&\partial _{\theta }\rho _c\left( \chi ,\theta \right) \dot{\theta}+\partial
_{\chi }\rho _c\left(\chi ,\theta \right) \dot{\chi }+\nonumber\\
&&3\frac{\dot{a}}{a}\left[ \rho _c
\left( \chi ,\theta \right) +\frac{p_c\left( \chi,\theta \right) }{c^{2}%
}\right] =\Gamma ,
\eea
where $\partial _{\theta }=\partial /\partial \theta $ and $\partial
_{\chi }=\partial /\partial \chi $, respectively.

In the present Subsection we consider the cosmological evolution of the condensate dark matter component only, by ignoring the effects of the radiation and of the baryonic matter. In order to simplify the formalism we introduce the critical density of the Universe as $\rho _{cr}=H_{0}^{2}/8\pi G$, where $%
H_{0}$ is the value of the Hubble function at $a=a_{0}$, and we  define the
density parameter of the dark matter as $\Omega _{DM,tr}=\rho _{tr}/\rho _{cr}$. We also rescale the comoving time variable $t$  by introducing the dimensionless time variable
\be
\tau =H_{0}\sqrt{\Omega _{DM,tr}}t,
\ee
respectively.
Accordingly, the constant $K$ in the expression of $\Gamma $ is rescaled as $\tilde{K}=K/H_{0}\sqrt{\Omega _{DM,tr}}$, and we denote the corresponding energy density transition rate by $\tilde{\Gamma }$. Moreover, we normalize the scale factor so that $a_0=1$ gives the value of the scale factor corresponding to the present age of the Universe.

By solving Eqs.~(\ref{eq1}) and (\ref{eq2}) for $\dot{\chi }$ and $\dot{\theta }$, we obtain the following differential equations describing the time variation of the temperature and of the condensate fraction,
\bea\label{T}
\frac{d \chi }{d\tau }&=&-\frac{3}{\Delta }\frac{1}{a}\frac{da}{d\tau }\left[ \left( \tilde{\rho}+\frac{%
\tilde{p}}{c^{2}}\right) \partial _{\theta }\rho _{c}-\left( \rho _{c}+\frac{%
p_{c}}{c^{2}}\right) \partial _{\theta }\tilde{\rho}\right] - \nonumber\\
&& \frac{\tilde{\Gamma }} {\Delta
}\left( \partial _{\theta }\rho _{c}+\partial _{\theta }\tilde{\rho}\right) ,
\eea
and
\be\label{Q}
\frac{d\theta }{d\tau }=-\frac{3}{\Delta }\frac{1}{a}\frac{da}{d\tau }
 \left( \rho _{c}+\frac{%
p_{c}}{c^{2}}\right) \partial _{\chi }\tilde{\rho} +
\frac{\tilde {\Gamma }}{\Delta }%
 \partial _{\chi }\tilde{\rho} ,
\ee
respectively, where we have denoted $\Delta =\partial _{\theta }\rho _{c}\partial _{\chi }%
\tilde{\rho}-\partial _{\chi }\rho _{c}\partial _{\theta }\tilde{\rho}=\partial _{\theta }\rho _{c}\partial _{\chi }%
\tilde{\rho}$, and we have taken into account that $\rho _c$ is independent of the reduced temperature $\chi $, $\partial _{\chi }\rho _c=0$.

The evolution of the scale factor of the Universe is described by the equation
\be\label{a}
\frac{1}{a}\frac{da}{d\tau }=\sqrt{\frac{\rho _{DM}\left( \chi,\theta \right) }{%
\rho _{tr}}},
\ee

Eqs.~(\ref{T}), (\ref{Q}) and (\ref{a}) give a complete description of the cosmological evolution of the finite temperature Bose-Einstein condensed dark matter. In order to completely determine the cosmological dynamics of the dark matter, we need the initial value of $\theta $ and $T$ at $a=a_{tr}$, $\theta \left(a_{tr}\right)=\theta _0$. Since at the initial moment of the transition there is no dark matter in the form of a condensate, it follows that $\theta _0=0$ at $a=a_{tr}$. As for the initial value of the reduced temperature, it is given by $\chi \left(a_{tr}\right)=1$.
In the following we denote
\be
\kappa _0=\frac{k_BT_{tr}}{mc^2}.
\ee

The cosmological evolution of the Bose-Einstein condensed dark matter depends on two dimensionless parameters, $\kappa _0$ and $\kappa $. By assuming for the condensation temperature $T_{tr}$ a value of the order of $T_{tr}=10^{10}$ K \cite{Har}, for a particle mass of the order of $m=10^{-37}$ g, we obtain $\kappa _0\approx 1.533\times 10^{10}$. The transition density corresponding to this temperature is $\rho _{tr}=7.255\times 10^{-22}$ g/cm$^3$, giving, for $l_a=10^{-20}$ cm, $\kappa =5.163\times 10^{14}$.  For a transition density of the order of $\rho _{tr}=10^{-9}$ g/cm$^3$, and for $l_a=10^{-20}$ cm, we obtain $\kappa _0=1.897\times 10^{18}$ and $\kappa =4.641\times 10^{10}$. However, due to the poor knowledge of the transition temperature and density, the values of these parameters are very uncertain. That's why in the following we will consider a wide range of values for the parameters $\kappa _0$ and $\kappa $, ranging from small to high values. The constants $\kappa _0$ and $\kappa $ are not independent, but they are related by the relation
\bea
\kappa _0&=&\frac{2\pi \hbar ^2}{\zeta ^{2/3}(3/2)c^2}\frac{1}{m^2l_a^2}\frac{1}{\kappa ^2}=\nonumber\\
&&4.088\times 10^{39}\left(\frac{m}{10^{-37}\;{\rm g}}\right)^{-2}\left(\frac{l_a}{10^{-20}\;{\rm cm}}\right)^{-2}\frac{1}{\kappa ^2}.\nonumber\\
\eea

Once the values of $\kappa $ and $\kappa _0$ are fixed, or obtained, for example, from general physical/observational considerations, the value of the product $ml_a$ is determined as
\be
ml_a=\frac{\sqrt{2\pi}}{\zeta ^{1/3}(3/2)}\frac{\hbar}{c}\frac{1}{\kappa \sqrt{\kappa _0}}=\frac{6.394\times 10^{-38}}{\kappa \sqrt{\kappa _0}}\;{\rm g\;cm}.
\ee
In the following we also denote
\be
K_1 = \widetilde{K}/l_a.
\ee

The time variation of the dimensionless condensate density is represented in Fig.~\ref{FIG5}.

\begin{figure}[!ht]
\includegraphics[height=60mm, width=0.98\linewidth]{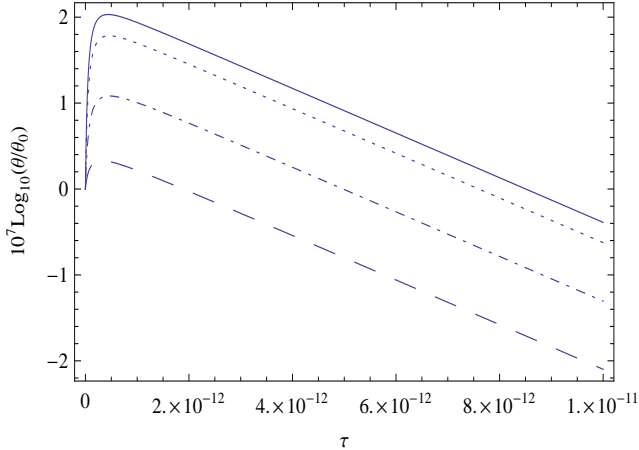}
\caption{Time variation of the dimensionless condensate density $\theta$ for $K_1 = 5 \times 10^{37}$  (solid curve), $K_1 =4\times 10^{37}$ (dotted curve),  $K_1 = 2\times 10^{37}$ (dot dashed curve), and $K_1 = 0.6 \times 10^{37}$ (dashed curve), respectively. The numerical values of the initial conditions and of the parameters are $\theta _0 = 10^{-9}$, $a_{tr}=10^{-4}$, $\chi _0 = 0.9999$, $\kappa = 10^{15}$ and $\kappa _0 = 10^{10}$.}
\label{FIG5}
\end{figure}

Due to the decrease of the temperature during the cosmological evolution, the condensate density increases immediately after the condensation phase transition, and it reaches a maximum value $\theta _{\max}$ at $\tau =\tau _{\max}$. But for time intervals so that $\tau >\tau _{\max}$, the cosmological expansion takes over, and  the density of the Bose-Einstein condensate dark matter decreases in time. The variation of the temperature of the  condensate dark matter after the start of the phase transition is represented in Fig.~\ref{FIG6}.

\begin{figure}[!ht]
\includegraphics[height=60mm, width=0.98\linewidth]{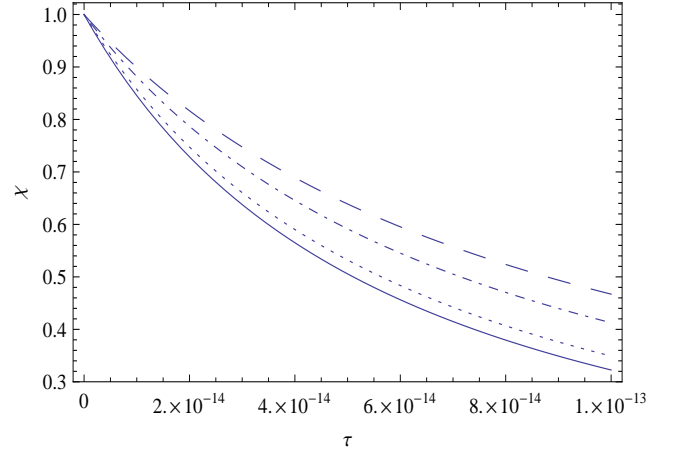}
\caption{Time variation of the reduced temperature $\chi =T/T_{tr}$ of the finite temperature condensed dark matter, for $K_1 = 5 \times 10^{37}$  (solid curve), $K_1 =4\times 10^{37}$ (dotted curve),  $K_1 = 2\times 10^{37}$ (dot dashed curve), and $K_1 = 0.6 \times 10^{37}$ (dashed curve), respectively. The numerical values of the initial conditions and of the parameters are $\theta _0 = 10^{-9}$, $a_{tr}=10^{-4}$, $\chi _0 = 0.9999$, $\kappa = 10^{15}$ and $\kappa _0 = 10^{10}$.}
\label{FIG6}
\end{figure}

The time variation of the total density of the dark matter, as well as of its total pressure, are represented in Figs.~\ref{FIG7} and \ref{FIG8}, respectively.

\begin{figure}[!ht]
\includegraphics[height=60mm, width=0.98\linewidth]{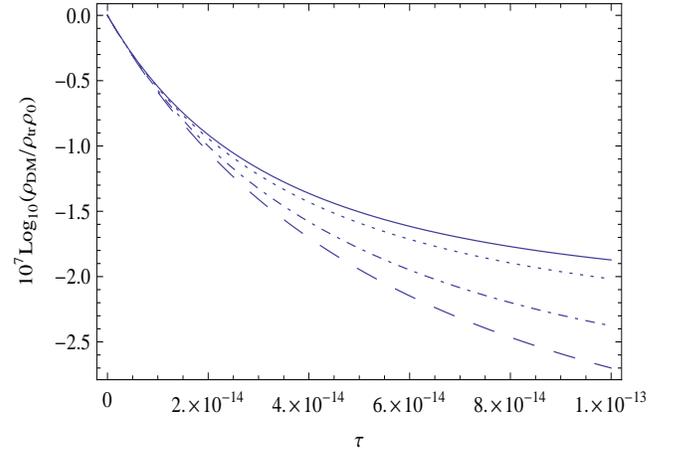}
\caption{Time variation of the total density $\rho _{DM}$ of the finite temperature condensed dark matter, for $K_1 = 5 \times 10^{37}$  (solid curve), $K_1 =4\times 10^{37}$ (dotted curve),  $K_1 = 2\times 10^{37}$ (dot dashed curve), and $K_1 = 0.6 \times 10^{37}$ (dashed curve), respectively. The numerical values of the initial conditions and of the parameters are $\theta _0 = 10^{-9}$, $a_{tr}=10^{-4}$, $\chi _0 = 0.9999$, $\kappa = 10^{15}$ and $\kappa _0 = 10^{10}$.  $\rho _0$ is the value of the condensate density corresponding to $\theta _0$ and $\chi _0$.}
\label{FIG7}
\end{figure}

\begin{figure}[!ht]
\includegraphics[height=60mm, width=0.98\linewidth]{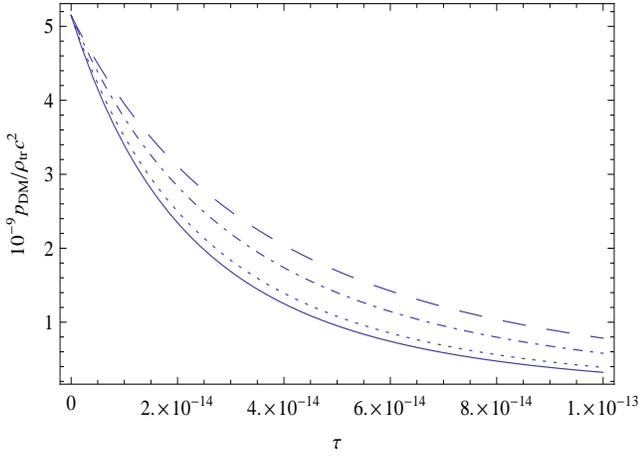}
\caption{Time variation of the total pressure $p_{DM}$ of the finite temperature condensed dark matter, for $K_1 = 5 \times 10^{37}$  (solid curve), $K_1 =4\times 10^{37}$ (dotted curve),  $K_1 = 2\times 10^{37}$ (dot dashed curve), and $K_1 = 0.6 \times 10^{37}$ (dashed curve), respectively. The numerical values of the initial conditions and of the parameters are $\theta _0 = 10^{-9}$, $a_{tr}=10^{-4}$, $\chi _0 = 0.9999$, $\kappa = 10^{15}$ and $\kappa _0 = 10^{10}$.}
\label{FIG8}
\end{figure}

We compare the cosmological evolution of the finite temperature condensed dark matter with the standard $\Lambda $CDM model of the pressureless dark matter, with total density $\rho _{DM}^{(nc)}$, and with the cosmological expansion described by the scale factor $a^{(nc)}$. From Eqs.~(\ref{dH}) and (\ref{drho}) we obtain immediately
\be
\frac{\rho _{DM}^{(nc)}}{\rho _{tr}}=\left(\frac{a_{tr}}{a}\right)^3,
\ee
and
\be
a^{(nc)}(\tau)=\left(\frac{3}{2}\right)^{2/3}\left(\tau +\frac{2}{3}a_{tr}^{3/2}\right)^{2/3},
\ee
respectively. The comparison of the time variation of the scale factor of the  Universe filled with condensed dark matter  with the standard pressureless $\Lambda $CDM dark matter model is represented in Fig.~\ref{FIG9}.

\begin{figure}[!ht]
\includegraphics[height=60mm, width=0.98\linewidth]{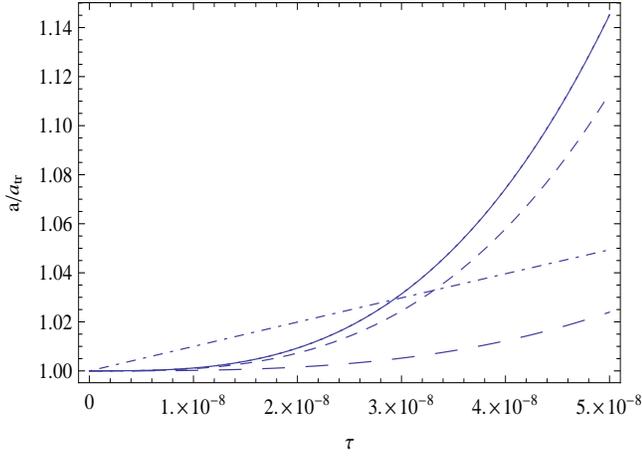}
\caption{Time variation of the scale factor of the finite temperature condensed dark matter for  $K_1 = 5 \times 10^{37}$  (solid curve), $K_1 = 4\times10^{37}$ (dotted curve), $K_1 = 2\times 10^{37}$ (short-dashed curve), $K_1 = 0.6 \times 10^{37}$ (long-dashed curve), and of the pressureless dark matter $\Lambda $CDM model  (dot-dashed). The numerical values of the initial conditions and of the physical parameters are  $\theta _0 = 10^{-9}$, $a_{tr}=10^{-4}$, $\chi _0 = 0.9999$, $\kappa = 10^{15}$ and $\kappa _0 = 10^{10}$.}
\label{FIG9}
\end{figure}

As one can see from the Figure, in the presence of the condensed dark matter the expansion of the Universe is faster.

\subsection{Cosmological evolution of the Universe with BEC dark matter, dark energy and radiation}

In the case of a Universe filled with dark energy, radiation,
baryonic matter with negligible pressure, and Bose-Einstein condensed dark matter, respectively, the time evolution
of the scale factor is given by the differential equation
\begin{widetext}
\begin{equation}\label{59}
\frac{1}{ a }\frac{d a }{dt}=H_{0}%
\sqrt{\frac{\Omega _{b,0}}{ a ^{3}}+\frac{\Omega _{rad,0}%
}{ a ^{4}}+\Omega _{DM,tr}\frac{\rho _{DM}}{\rho _{tr}}%
+\Omega _{\Lambda }},t\geq t_{tr},
\end{equation}
\end{widetext}
which must be integrated with the initial condition $a\left(t_{tr}\right) =a_{tr}$. In the following for the Hubble constant we adopt the value $H_0 = 70\; {\rm km/s/Mpc} = 2.273 \times 10^{-18}$ s$^{-1}$ \cite{Hin}, giving for the critical density a value of $\rho _{c,0} = 9.248 \times 10 ^{-30}$ g/cm$^3$. For $\Omega _{DM,0}$  $\Omega _{b,0}$, $\Omega _{rad,0}$, and $\Omega _{\Lambda }$ we adopt the numerical values $\Omega _{DM,0}=0.228$, $\Omega _{b,0} = 0.0456$, $\Omega _{rad,0} = 8.24 \times 10^{-5}$, and $\Omega _{\Lambda } = 0.726$ \cite{Hin}, respectively. By taking into account that $\rho _{tr}a_{tr}^3\approx \rho _{c,0}$, it follows that $\Omega _{DM,tr}\approx\Omega _{DM,0}/a_{tr}^3$, and the evolution equation of the scale factor can be written as
\begin{equation}
\frac{1}{a}\frac{da}{d\tau }=\sqrt{\frac{\Omega _{b,0}}{\Omega _{DM,tr}}%
\frac{1}{a^{3}}+\frac{\Omega _{rad,0}}{\Omega _{DM,tr}}\frac{1}{a^{4}}+\frac{%
\rho _{DM}}{\rho _{tr}}+\frac{\Omega _{\Lambda }}{\Omega _{DM,tr}}}.
\end{equation}

The comparison of the time variation of the scale factor of the  Universe filled with dark energy, radiation,
baryonic matter with negligible pressure, and Bose-Einstein condensed dark matter, respectively, and with the standard pressureless $\Lambda $CDM dark matter is represented in Fig.~\ref{FIG10}.

\begin{figure}[!ht]
\includegraphics[height=60mm, width=0.98\linewidth]{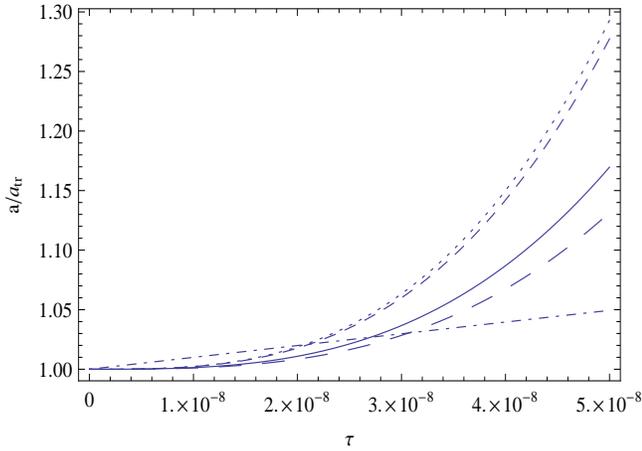}
\caption{Time variation of the scale factor of the Universe filled with dark energy, radiation,
baryonic matter with negligible pressure, and Bose-Einstein condensed dark matter for $K_1 = 4 \times 10^{36}$  (solid curve), $K_1 = 3 \times 10^{36}$ (dotted curve), $K_1 = 2\times 10^{36}$ (short-dashed curve), $K_1 = 0.6 \times 10^{36}$ (long-dashed curve), and of the standard  $\Lambda $CDM model (dot-dashed curve). The numerical values of the initial conditions and of the physical parameters are $\theta _0 = 10^{-9}$, $a_{tr}=10^{-4}$, $\chi _0 = 0.9999$, $\kappa = 10^{15}$ and $\kappa _0 = 10^{10}$.}
\label{FIG10}
\end{figure}

\section{Discussions and final remarks}\label{5}

In the present paper we have considered the cosmological dynamics of the finite temperature Bose - Einstein Condensed dark matter. In a simplified approach one can describe a finite temperature condensed system as a mixture of two fluids in thermodynamic equilibrium, the thermal cloud, consisting of the thermal excitations of the condensate, and of the proper condensate state. At the beginning of the condensation process the bosonic dark matter is mostly in the form of the thermal cloud. However, once the temperature of the system drops, more and more particles from the thermal cloud will enter into the condensate energy band. Therefore, the cooling of the Universe due to the cosmological expansion did lead to the significant decrease in the number of particles in the thermal cloud, and to the corresponding increase in the number of the condensate particles. We have described the cosmological transition process in the framework of a two interacting fluid model. Due to the cooling of the Universe, the thermal cloud looses its particles and the thermal excitations enter into the condensate. Therefore, the dynamics of the dark matter is determined by the increase of the density $\rho _c$ of the condensed component, which is initially increasing in time, and the corresponding decrease in the number of the thermal particles. However, after reaching a maximum value, the density of the condensate component will start to decrease. This happens at the moment when the time rate of the decrease in the condensate density due to the expansion of the Universe becomes higher than the density transfer rate from the thermal cloud. As one can see from Eq.~(\ref{Q}) the maximum value of the condensate density is reached when $d\theta /d\tau =0$, which gives the algebraic equation
\be \label{gamma}
\left.\tilde {\Gamma } \right|_{\theta =\theta _{max}}=\left.3\sqrt{\frac{\rho _{DM}\left( \chi,\theta \right) }{%
\rho _{tr}}} \left( \rho _{c}+\frac{p_{c}}{c^{2}}\right)\right|_{\theta =\theta _{max}} ,
\ee
for the determination of the numerical values of $\theta _{max}$ as a function of the reduced temperature $\chi $ and of the physical parameters of the condensed dark matter system.  The contour plot of Eq.~(\ref{gamma}) is represented in Fig.~\ref{FIG11}. A contour plot is a plot of equipotential curves $z(x,y)$, where each of the curve is the geometric locus of pairs $\{ x,y\}$ such that $z = {\rm constant}$. In the case discussed here, each curve tracks the set of pairs $\{ \theta _{max},\chi \}$ for which the value of the right hand side of Eq.~(\ref{gamma}) is a constant (the values of these constants are shown explicitly on each curve).

\begin{figure}[!ht]
\includegraphics[height=60mm, width=0.98\linewidth]{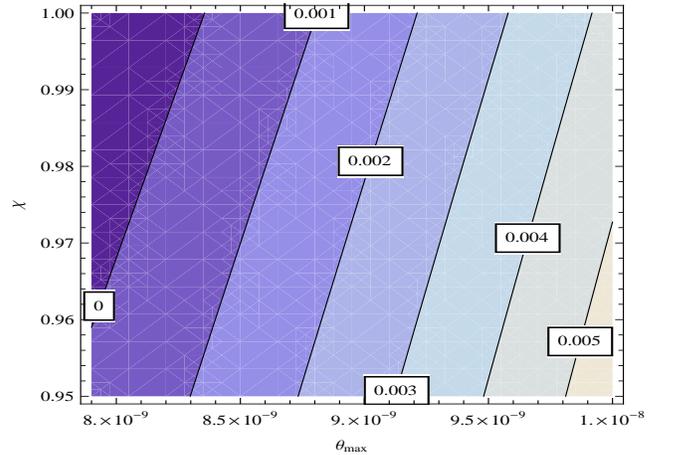}
\caption{The contour plot of Eq.~(\ref{gamma}), for $K_1 = 0.6 \times 10^{37}$, $\kappa = 10^{15}$ and $\kappa _0 = 10^{10}$.}
\label{FIG11}
\end{figure}

As a general result we have also found that the presence of the condensed dark matter will accelerate the expansion of the Universe. In the case of zero temperature condensed dark matter this behavior has been already pointed out \cite{mnr,chav, Har}. However, a further increase in the speed of the expansion may be expected due to the finite temperature effects during the phase transition. Moreover, due to the presence of the two interacting fluid components, the cosmological dynamics of the finite temperature condensed dark matter is much more complicated than expected. Therefore the two-fluid finite temperature condensation transition may provide some clear cosmological signatures that could help in discriminating between Bose - Einstein Condensed dark matter models, and the standard pressureless $\Lambda $CDM cosmological model.

\acknowledgments

TH is supported by an RGC grant of the government of the Hong Kong SAR. GM acknowledges the financial support of the Sectoral Operational Programme for Human Resources Development 2007-2013, co-financed by the European Social Fund, under the project number POSDRU/107/1.5/S/76841 with the title "Modern Doctoral Studies: Internationalization and
Interdisciplinarity". GM would like to thank the Institute for Theoretical Physics, Vienna University of
Technology, Austria, for their hospitality during the time when this work was drafted.

\end{document}